\documentclass[useAMS,usenatbib]{mn2e}

\usepackage{epsf}
\usepackage{amsmath}

\newcommand{\kmsmpc}{\kms\;{\rm Mpc}^{-1}}

\newcommand{\hkpc}{h^{-1}{\rm kpc}}
\newcommand{\hmpc}{h^{-1}{\rm Mpc}}

\newcommand{\kms}{\;{\rm km}\,{\rm s}^{-1}}

\newcommand{\gad}{{\sc Gadget-2}}
\newcommand{\ion}[2]{\hbox{#1\,{\sc #2}}}
\newcommand{\vw}{v_{\rm w}}

\title[Enrichment \& Pre-Heating in Groups from Outflows]{Enrichment and Pre-Heating in Intragroup Gas from Galactic Outflows}

\author[R. Dav\'e, B. D. Oppenheimer, S. Sivanandam]{Romeel Dav\'e, Benjamin D. Oppenheimer, Suresh Sivanandam\\Astronomy Department, University of Arizona, Tucson, AZ 85721}

\begin{document}

\pubyear{2008}

\maketitle

\label{firstpage}

 \begin{abstract}
We examine metal and entropy content in galaxy groups having
$T_X\approx0.5-2$~keV in cosmological hydrodynamic simulations.
Our simulations include a well-constrained prescription for galactic
outflows following momentum-driven wind scalings, and a sophisticated
chemical evolution model.  Our simulation with no outflows reproduces
observed iron abundances in X-ray emitting gas, but the oxygen abundance
is too low; including outflows yields iron and oxygen abundances in
good agreement with data.  X-ray measures of [O/Fe] primarily reflect
metal distribution mechanisms into hot gas, not the ratio of Type~Ia
to Type~II supernovae within the group.  Iron abundance increases by
$\sim\times2$ from $z\sim 1\rightarrow 0$ independent of group size,
consistent with that seen in clusters, while [O/Fe] drops by $\sim30$\%.
Core entropy versus temperature is elevated over self-similar predictions
regardless of outflows due to radiative cooling removing low-entropy gas,
but outflows provide an additional entropy contribution below 1~keV.
This results in a noticeable break in the $L_X-T_X$ relation below $\sim
1$~keV, as observed.  Entropy at $R_{\rm 500}$ is also in good agreement
with data, and is unaffected by outflows.  Importantly, outflows serve to
reduce the stellar content of groups to observed levels.  Specific energy
injection from outflows drops with group mass, and exceeds the thermal
energy for $\la 0.5$~keV systems.  Radial profiles from simulations
are in broad agreement with observations, but there remain non-trivial
discrepancies that may reflect an excess of late-time star formation
in central group galaxies in our simulations.  Our model with outflows
suggests a connection between physical processes of galaxy formation
and both pre-heating and enrichment in intragroup gas, though more
definitive conclusions must await a model that simultaneously suppresses 
cooling flows as observed.
\end{abstract}

 
\section{Introduction}

Galaxy clusters are the largest bound structures in the Universe.
This oft-repeated phrase captures both their extremeness as well as their
simplicity.  Clusters are virialized and nearly spherical, their baryons
are mostly in the form of a hot intracluster medium (ICM), and their
galaxies are nearly all early-types with old stellar populations.
The ICM is close to hydrostatic equilibrium with a temperature near
the halo's virial temperature, and is enriched to around one-third solar
independent of cluster mass.  The only significant subdivision of clusters
comes from examining their central gas properties, where they split into
cool core (or cooling flow) and non-cool core clusters.  Their smaller
counterparts, galaxy groups (usually referring to systems with
X-ray temperatures below 2~keV or so), share many of these characteristics.
The simplest models regard groups as mini-clusters, i.e. bound spheres
of hot gas in hydrostatic equilibrium, whose properties should be
self-similar with clusters~\citep[e.g.][]{kai86,nav95}.

But a closer inspection of group and cluster properties reveals many
puzzles and complexities.  For instance, gas in hydrostatic equilibrium
near the centers of groups should be sufficiently dense to have
cooling times significantly shorter than a Hubble time~\citep{fab84},
and therefore gas should be cooling rapidly onto the central galaxy.
Yet X-ray observations show a distinct lack of gas at temperatures
below roughly one-third of the system temperature, and little
cold gas or star formation is seen in the center~\citep[see][for
review]{pet06}.  Evidently, some source of energy or entropy is acting
to prevent gas cooling in the core.  Recently, observations of X-ray
bubbles in the ICM apparently blown by AGN-driven jets from the central
galaxy offer a possible way to balance cooling~\citep{mcn07,dun08}.
Cosmic rays~\citep{guo08,sij08}, conduction~\citep{kim03a,jub04}, and
gravitational heating~\citep{kim03b,bur08,dek08} may also be viable
mechanisms.

Another indication of non-gravitational energy injection into
intracluster gas comes from scaling relations between cluster X-ray
luminosity ($L_X$), X-ray luminosity-weighted temperature ($T_X$),
and galaxy velocity dispersion ($\sigma$).  The self-similar model
including only gravitational heating predicts $L_X\propto T_X^2$,
which is confirmed by hydrodynamic simulations that do not include gas
cooling or processes associated with galaxy formation~\citep{nav95,eke98}.
Observations indicate $L_X\propto T_X^3$ for clusters~\citep{whi97}, and
an even steeper relation in groups \citep[e.g.][]{xue00}.  This has led
to speculation that some feedback process has added a significant amount
of non-gravitational energy into the ICM, which would puff out the gas
distribution and lower the X-ray luminosity.  Such energy injection was
dubbed ``pre-heating" by \citet{kai91}. Understanding the amount, epoch,
and origin of pre-heating has been the subject of many investigations.

An influential paper by \citet{pon99} claimed that systems above 1--2~keV
showed rising entropy with temperature, while below that temperature all
groups had a similar entropy of $\sim 100$~keV cm$^2$.  Here entropy is
defined as the X-ray weighted temperature divided by the electron density
at one-tenth the virial radius to the two-thirds power.  While subsequent
observations showed that this claimed entropy floor is instead merely a
more-gradual-than-expected decline~\citep[e.g.][]{hel00}, the question of
what elevates the entropy of gas in smaller systems remains unanswered.
Entropy has now become the de facto quantity discussed when quantifying
pre-heating.

A third, more oblique indication of non-gravitational energy input
into the ICM is the presence of metals.  The amount of metals, roughly
one-third solar in clusters regardless of size, is large when one
considers that in the most massive clusters only a few percent of baryons
have formed into stars.  It is further remarkable that this enrichment
level is similar in clusters of all masses despite significant trends
with mass in the amount of stars formed~\citep[e.g.][]{gon07,balogh07}.
Enriching the ICM to these levels represents a non-trivial challenge,
because stars in cluster galaxies can themselves only provide a
metallicity of about one-sixth solar~\citep{por04}.  Accounting for
intracluster stars can alleviate this disrepancy~\citep{zar04,siv08}, but
the metals still must be removed from galaxies yet be retained in the hot
ICM gas.  One possibility is that gas stripped off of infalling galaxies
can sufficiently enrich the ICM~\citep{vank07}; if so, the enrichment
process would not be energetically important.  However, it is clear that
the diffuse intergalactic medium (IGM) is enriched at as early as $z\sim
6$ by powerful outflows~\citep{agu01,opp06}, so perhaps such outflows
could also be responsible for enriching intracluster and intragroup gas.
If so, outflows might simultaneously provide some level of pre-heating.

In short there are three current puzzles in understanding the physics of
the ICM:
\begin{itemize}
\item {\it Cooling flow problem:} Cooling rates of hundreds to thousands
of solar masses per year are expected based on simple cluster models,
yet the measured amount of cooling gas is at least an order of magnitude
less.
\item {\it Entropy problem:}  Entropy levels are progressively more
elevated over self-similar predictions towards smaller systems.
\item {\it Metallicity problem:}  It is unclear how metals are removed 
from galaxies in such a way as to enrich intracluster and intragroup gas
as observed.
\end{itemize}

Occam's razor leads one to attempt to solve these issues with a single
physical mechanism.  Unfortunately, this is difficult.  For instance,
the solution to the cooling flow puzzle cannot invoke significant star
formation (and hence metal production), because little recent star
formation is seen in cluster galaxies.  Models of entropy injection from
supernovae seem to suggest that it is insufficient to solve the entropy
problem~\citep{bal99,bow01,pip02}.  These days, it is fashionable to mostly
neglect the metallicity problem, ascribing it to a change in stellar
yields or perhaps even the initial mass function (IMF), and attempt to
solve the cooling flow and entropy problems with a single energy source:
heating from active galactic nuclei (AGN).  However, this has difficulties
as well, because the entropy requirements are large~\citep{bab02,mccar02},
though more recently analytic models that use AGN in a self-regulated
way have had some success~\citep{mccar08}.  

In this paper we focus on the entropy and metallicity problems,
leaving aside the cooling flow issue for now.  Previous theoretical
studies of the entropy problem have generally assumed some amount of
energy or entropy injection at an early epoch, and then attempted to
constrain the amount and epoch of pre-heating by evolving systems
(either analytically or numerically) to compare with present-day
clusters~\citep[e.g.][]{bab02,kay04,bor05}.  Few attempts have been
made to understand or model in detail possible sources of pre-heating.
Observationally, there have been attempts to connect pre-heating and metal
enrichment, but with limited success.  \citet{ren01} argued that galactic
winds must be responsible for ICM enrichment, but they could only add
a negligible amount of energy.  \citet[][hereafter PSF03]{pon03} found
that neither isentropic pre-heating nor cooling models can reproduce the
entropy structure of intragroup gas.  Hence the physics of pre-heating
remains a mystery, and the constraints on its amount and epoch remain
only loosely connected to physical process of galaxy formation.

Here we take a different approach.  We begin with a model for galactic
outflows within hierarchical galaxy formation simulations that is
constrained by non-ICM related observations, such as high-redshift IGM
enrichment and the galaxy mass-metallicity relation.  We then study how
such outflows impact the enrichment and entropy level of intragroup
gas at $z=0$.  We are limited to investigating galaxy groups and not
clusters owing to our numerical capabilities, but this is actually
desirable because groups fall in a particularly interesting regime for
pre-heating and enrichment processes.  The observed turn-down at $\sim
1$~keV in the $L_X-T_X$ scaling relations~\citep{xue00} indicates that
pre-heating energy input is comparable to the system's gravitational
energy at group scales.  Groups have a signficantly larger stellar
fraction, yet their metallicity appears to be comparable to or even lower
than clusters, indicating that metal transport processes are crucial.
Groups are more generally important because they contain the majority
of galaxies in the Universe, and simple dynamical arguments suggest
that group environments are the most effective for morphologically
transforming galaxies through mergers.  Hence there appears to be a
lot of action at the group scale.  Observations of intragroup gas are
rapidly improving, and this provides important constraints on models of
pre-heating and enrichment.  Investigating such constraints is a central
theme of this paper.

We have recently identified a model for galactic outflows that
is remarkably successful at reproducing a wide range of observations
of cosmic chemical enrichment.  In \citet[][hereafter OD06]{opp06}
we used cosmological simulations that incorporate enriched kinetic
feedback to show that a certain class of outflow models are particularly
successful at matching detailed properties of IGM
enrichment, as traced by \ion{C}{iv} absorbers in quasar spectra from
$z\sim 6\rightarrow 2$.  In \citet{fin08} we showed that the exact same
model was concurrently and fairly uniquely successful at reproducing
the observed mass-metallicity relation in galaxies.  In \citet{dav06} we
showed that such winds are able to reproduce the luminosity functions of
early galaxies, significantly suppressing early star formation as needed
to solve the overcooling problem.  \citet{bou07} showed that such a
model best reproduces the early cosmic luminosity density evolution.
The success of a single outflow model in matching such a wide range
of observations, albeit mostly at $z\ga 2$, is compelling, and merits
investigation into its impact on intragroup and intracluster gas.

The outflow model we have found to be most successful follows scalings
predicted by a momentum-driven wind scenario~\citep{mur05}.  In such
a model, the wind speed scales with the galaxy's circular velocity,
while the mass loading factor, i.e. the amount of mass ejected in an
outflow as compared to the star formation rate, scales inversely with it.
This yields a constant momentum input per unit star formation.  We have
found that the generic property making this outflow model successful
is that early galaxies drive out an amount of mass comparable to or
exceeding that formed into stars, at speeds well above the escape velocity
of their host halos.  While this might seem suprising, it appears to be
required for early IGM enrichment~\citep{agu01}, global star formation
suppression~\citep{spr03b}, and establishing the gas and metal content
of high-$z$ galaxies~\citep{erb08}.  Intriguingly, this scaling of wind
velocity with galaxy size is also observed in local starburst-driven
outflows~\citep{mar05,rup05}, and at $z\sim 1$ in star-forming
galaxies~\citep{wei08}, providing a connection between rare present-day
outflows and the ubiquitous ones at high-$z$~\citep[e.g.][]{ste04}.
One notable issue is that for sufficient mass to be ejected, the
energetic requirements are large, comparable to the available supernova
energy~\citep{opp08}.  Given expected radiative losses, it may be that
another source of energy besides supernovae is required, such as photons
from young stars in low-mass galaxies, or AGN in high-mass galaxies.
The successes of these momentum-driven wind scalings suggest that we
now have a plausible model, albeit phenomenological, of how outflows
move mass, metals, and energy on large scales within a hierarchical
structure formation scenario.

In this paper we show that cosmological hydrodynamic simulations
incorporating momentum-driven outflow scalings are able to simultaneously
enrich intragroup media, reduce the stellar fraction to observed levels,
and match the entropy levels of observed groups.  We compare simulations
with and without winds to assess the impact of outflow metal and
energy injection.  We find that radiative cooling is mainly responsible
for breaking the self-similarity between groups and clusters, but that
outflows provide an additional entropy boost that becomes important in the
poorest groups ($T_X\la 1$~keV).  A key aspect of outflows is that they
suppress star formation, because cooling-only models can match intragroup
medium entropy levels only at the expense of excessive stellar content.
Our results suggest that pre-heating and enrichment in groups today
arise from the same outflows that are required to distribute metals in
the cosmos and regulate early star formation.

In \S\ref{sec:sims} we describe our Gadget-2 simulations with our
chemical evolution and outflow modeling.  In \S\ref{sec:enrichment}
we study the enrichment properties of intragroup gas in these models,
including alpha element ratios and evolution from $z=1\rightarrow 0$.
In \S\ref{sec:entropy} we examine entropy levels and scaling relations
in intragroup gas, and quantify energy injection from outflows.
In \S\ref{sec:profiles} we compare radial profiles of physical properties
to data, and study the impact of outflows on such profiles.  We summarize
in \S\ref{sec:summary}.

\section{Simulations}\label{sec:sims}

\subsection{Simulations with outflows}

We run cosmological hydrodynamic simulations using \gad~\citep{spr05},
a cosmological Tree-Particle Mesh-Smoothed Particle Hydrodynamics
code including radiative cooling and star formation~\citep{spr03a}.
Our version~\citep[hereafter OD08]{opp08} also includes metal-line
cooling~\citep{sut93}, uses a (spatially-uniform) cosmic photoionizing
background given by~\citet{haa01}.

We include the effects of galactic outflows and the creation of heavy
elements as described in OD08.  This builds on our implementation
detailed in \citet{opp06}, following \citet{spr03a}, with three
major improvements: (1) We track enrichment contributions from Type~II
supernovae, Type~Ia supernovae, and mass and metal loss from AGB stars;
(2) We individual follow carbon, oxygen, silicon, and iron enrichment;
(3) We use galaxy baryonic masses (rather than the local gravitational
potential) to estimate the galaxy parameters from which outflow properties
are derived, using an on-the-fly galaxy finder.  The upshot of these
improvements is a more accurate tracking of metal evolution in various
phases down to $z=0$, as well as a more realistic implementation of the
dependence of outflow properties with galaxy size.  Our chemodynamical
model is broadly similar to that used to study ICM enrichment using \gad\
by \citet{tor07}, though some details differ.  We employ \citet{bru03}
models with a Chabrier IMF for computing stellar evolution.

The differences in our new implementation versus the one in OD06 are
motivated primarily by a desire to treat the momentum-driven wind model
more accurately.  In our old model, we used the local gravitational
potential as a proxy for galaxy velocity dispersion.  At high-$z$,
when the dominant star-forming galaxy is the central object and halos
generally only contain a single large galaxy~\citep[i.e. the halo
occupation distribution lies on the plateau portion; e.g.][]{ber03}, this
is a reasonable assumption.  At low-$z$, as halos grow, satellites become
more numerous and house more of the total star formation, this becomes a
poorer assumption.  A galaxy driving an outflow does so locally, and does
not necessarily know about the halo in which it resides.  Hence in OD08 we
moved to identifying individual galaxies and using their properties rather
than the global halo properties to derive wind parameters.  In OD08 we
presented a detailed comparison of the resulting global and IGM enrichment
properties between our implementations, and found that at high-$z$ they
gave similar results, but at low-$z$ they differed.  We believe our new
implementation is more accurate particularly at $z=0$, so we employ it
here, and do not compare to our previous (OD06) implementation.

A key aspect for studies of the ICM is iron enrichment from Type~Ia
supernovae, included in our new simulations.  We employ a Type~Ia rate
following \citet{sca05}, who parameterize observations by \citet{man05}
via a prompt component that tracks star formation plus a delayed component
that tracks stellar mass.  The formulae and metal yields are detailed
in OD08.  Metals and energy from delayed Type~Ia SNe (coming from stars)
are added to the nearest three gas particles.  As noted in OD08, the
vast majority of iron in the universe is produced in Type~II supernovae,
but for concentrated stellar systems like groups and clusters, Type~Ia's
provide a large contribution to the gas-phase iron abundance~\citep[see
also][]{tor07}.

As in \citet{spr03a}, each particle that is eligible for star formation
has its star formation rate computed based on a \citet{ken98} law, and
an appropriate level of enrichment is added based on Type~II supernova
yields (see OD08).  The star-forming particle has some probability
to be kicked into an outflow or spawn a star particle.  The ratio
of those probabilities is given by the mass loading factor $\eta$,
which is an input parameter to the model.  If a particle is selected
to be in an outflow, it is kicked with a velocity $\vw$ in a direction
given by $\pm${\bf v}$\times${\bf a} (velocity cross acceleration),
resulting in a quasi-bipolar outflow\footnote{Movies can be seen at
{\tt http://luca.as.arizona.edu/\~{}oppen/IGM/}.}.  Wind particles
propagate purely gravitationally until they are outside the star-forming
region (typically few kpc), at which point they can again interact
hydrodynamically with surrounding gas.  The following formulae are used
to relate the galaxy baryonic mass $M_b$ to the mass loading factor and
wind speed for star-forming particles in that galaxy:
\begin{eqnarray}
\sigma &=& 200 \left(\frac{M_{b}}{5\times10^{12}M_\odot} \frac{\Omega_m}{\Omega_b} h\frac{H(z)}{H_{0}}\right)^{1/3} \kms,\\
\vw &=& 4.3 \sigma \sqrt{f_L-1},\\ \label{eqn:vwind}
\eta &=& \frac{150 {\rm km/s}}{\sigma},
\end{eqnarray}
where the first equation is taken from \citet{mur05}, originally based
on \citet{mo98}.  The value of $f_L$ is taken from observations of
outflows by \citet{rup05}, who found that they lie between $1.05-2$;
we choose a random number in that interval for every outflow event.
We include a factor to account for increased photon production in
lower-metallicity systems.  This wind model is identical in form to the
``vzw" model of \citet{opp06}; full details are available in OD08.

Ejected wind particles are hydrodynamically decoupled (i.e. the
hydro forces are turned off, but not gravity) until they reach a
density one-tenth of the critical density for star formation, or for
a maximum period of $20\hkpc/v_w$, which is e.g.  $\approx 50$~Myr for
$v_w=500$~km/s.  This is intended to simulate the outflow blowing a hole
through the interstellar medium, which cannot be accurately represented
given the smoothed nature of SPH.  \citet{dal08} have investigated
the effects of turning on or off hydrodynamic decoupling, and found
that it gives significantly different results for high-resolution
simulations of disk galaxies.  However, we have done similar tests at
the cosmological resolution used here, and found little difference in
the global enrichment properties; this is perhaps not surprising since
the spatial resolution here is $5\hkpc$, and the maximum distance an
outflow particle can propagate during the decoupled phase is $\approx
20\hkpc$ (and typically it is much less).  One advantage of decoupling,
as pointed out by \citet{dal08}, is that it results in better resolution
convergence.  Of course this may not be desireable if the result
converges to a physically incorrect answer, but it at least allows a
resolution-independent calibration of the outflow model.

We consider two main runs in this paper.  They are identical in every way
(including random phases in the initial conditions) with the exception
that one includes momentum-driven outflows as described above, while
the other does not include outflows.  Note that in both cases, thermal
energy feedback from supernovae is still accounted for in the multi-phase
subgrid model of \citet{spr03a}; however, the explicitly coupled phases
means that such feedback cannot drive an outflow.  This is the reason
why an explicit outflow model is required~\citep{spr03b}.

Each run has $256^3$ dark matter and $256^3$ gas particles in a
randomly-generated volume of $(64\hmpc)^3$ (comoving), with a concordance
cosmology~\citep{kom08} having $\Omega=0.25$, $\Lambda=0.75$, $n=0.95$,
$H_0=70\kmsmpc$, $\sigma_8=0.83$, and $\Omega_b=0.044$.  The value of
$\sigma_8$ is slightly higher than that favored from WMAP-5, moving more
towards values derived from cluster abundances~\citep[e.g.][]{evr08}; the
difference is not important for the present work.  Initial conditions
are generated using an \citet{eis99} power spectrum in the linear
regime at $z=129$ by displacing particles off a grid according to the
Zel'dovich approximation.  Particle masses are $2.72\times 10^8M_\odot$
and $1.27\times 10^9M_\odot$ for gas and dark matter, respectively,
and the gravitational softening length is set to $5\hkpc$ comoving.

\subsection{Computing group properties}

To identify galaxy groups, we find large bound halos using a spherical
overdensity (SO) algorithm, as described in \citet{ker05}.  This algorithm
finds particles at local potential minima, then expands spherically
around those particles until the mean overdensity enclosed corresponds to
the virial overdensity for the assumed cosmology \citep[specifically,
$178\Omega^{0.45}\rho_{\rm crit}=95\rho_{\rm crit}$;][]{eke98}.
The outskirts of these systems may not properly follow the ellipticity of
real systems (which is often ambiguous anyway), but since the outermost
properties contribute little to global group properties particularly when
weighted by X-ray emission, this difference is not likely to be important.

To identify galaxies, we use Spline Kernel Interpolative Denmax
\citep[SKID;][]{ker05}.  Groups of galaxies are taken to be SO halos that
contain $\geq5$ resolved galaxies.  Our galaxy mass resolution limit is
given by a stellar mass of $M_*>8.72\times 10^9 M_\odot$ (i.e. 64
star particle masses), above which galaxy stellar properties have been
shown to be converged~\citep{fin06}.  For comparison, this mass is
$\approx 0.2 M^*_{\rm star}$~\citep{bal08}, where $M^*_{\rm star}$ is
the characteristic mass in a Schechter function fit to the present-day
stellar mass function.

The X-ray luminosity of individual gas particles is computed
using the Astrophysical Plasma Emission Code\footnote{\tt
http://cxc.harvard.edu/atomdb/sources\_apec.html} (APEC)
models~\citep{smi01}, from the particle's mass, SPH-weighted density,
temperature, and metallicity.  APEC outputs X-ray spectra, from which
the luminosity is computed by summing intensities from 0.5--10~keV,
uniformly weighted (i.e. no filter profile is assumed).  Restricting to
0.5--2~keV makes little difference for our systems.  Contributions to
line and continuum emission from individually tracked elements (iron,
oxygen, silicon, and carbon, along with hydrogen and helium) are computed
separately and summed.  Other elements are tied to iron assuming solar
abundance ratios~\citep{and89}, though because line cooling is dominated
by iron and oxygen in the intragroup gas temperature and density regime,
this choice makes little difference.

The X-ray luminosity $L_X$ of a given group is the summed X-ray
luminosities of all particles within the virial radius.  The group
temperature $T_X$ is computed as the X-ray luminosity-weighted
temperature of gas particles, and all quoted metallicities are also X-ray
emission-weighted.  The emission weighting is done to mimic observations,
which obtain temperatures and abundances by fitting APEC or similar models
to X-ray spectra, and hence are effectively measuring luminosity-weighted
quantities.  Radial profiles are computed by performing similar summations
in 25 equal thickness spherical shells about the SO halo center.

\subsection{Resolution convergence}

In this paper we will not directly present any numerical resolution convergence
tests.  This is unfortunate, and is not our preferred mode of operation.
However, several practical issues limit our ability to directly study
convergence for the objects of interest, primarily because the groups
we examine in this work are already at the extreme of what we can
compute in a random-volume simulations.  While we have smaller-volume
simulations available, they do not produce any systems in the observed
group-mass regime, so direct convergence tests on group properties are
not possible.  To be specific, our $64\hmpc$ run's largest group is 2~keV,
which is already fairly small by X-ray group standards; our $32\hmpc$
run's largest is around 0.5~keV, which then does not allow us to check
resolution effects for the observationally relevant $0.5-2$~keV regime.
On the flip side, we cannot run even larger volume simulations given our
present computational capabilities, because worsening the mass resolution
beyond our $64\hmpc$ run is found to significantly compromise early
galaxy formation, which is important because these large bound systems
form much of their galaxy population early on.  Hence, we are stuck in
the uncomfortable position of only having a single combination of mass
resolution and volume that can examine the properties relevant for
observed galaxy groups.  For these reasons, the results in this work
must be considered preliminary, and our main goal is to present new
physical interpretations for group scaling relations as plausible if
not definitive.

Nevertheless, we do have some reason to believe that our results are
robust.  In OD08 we present resolution convergence studies for cosmic
mass, energy, and metal distribution, and the simulation resolution
used here is reasonably well converged.  We have also checked using a
$32\hmpc$ simulation with $256^3$ dark matter and gas particles that
the group properties at temperatures below 0.5~keV are numerically
converged.  We hope to be able to run constrained simulations to form
individual groups and clusters at higher resolution in the future,
but for now we do not take that approach because we would like to
compare large statistical samples to observations.  However, such a
constrained-realization study for clusters was done by \citet{tor07}
also using \gad, with mass resolution comparable to what we use here,
and through direct tests they determined that their results are not
very sensitive to numerical resolution, at least in the central regions
of clusters that dominate the X-ray emission.  Hence we believe that
our results here should be sufficiently robust to numerical effects,
so that our conclusions should be generally correct.

\section{Metal Content of Intragroup Gas}\label{sec:enrichment}

Observations of cluster and group enrichment have been bolstered by
the increased sensitivity provided by {\it Chandra} and {\it XMM}.
It is well known that clusters show an iron abundance of one-third
to one-half solar (outside the cool core), essentially independent of
cluster size~\citep[e.g.][]{ren97,pet03,deg04}.  Measurements in the poor
group regime ($T_X\la 2$~keV), which have recently become available, show
a larger scatter with a tendency towards slightly increased metallicity
relative to rich clusters, although the poorest groups seem to reverse the
trend and have lower abundances~\citep{davis99,hel00}.  It is puzzling
that despite a significant increase in the stellar baryon fraction from
the rich clusters to groups~\citep[e.g.][]{gon07}, no corresponding
change is seen in the X-ray gas metallicity.  This suggests that the
intragroup/intracluster medium metal enrichment is not simply related
to the stars formed within the halo potential, and that distribution
mechanisms and sources of enrichment play a critical role in governing
metallicities.  Hence the enrichment level of intragroup gas provides
key constraints on such processes~\citep[see review by][]{bor08}.
In this section we compare our simulations with and without outflows to
observations of intragroup gas metallicities.

\begin{figure}
\vskip -0.5in
\setlength{\epsfxsize}{0.6\textwidth}
\centerline{\epsfbox{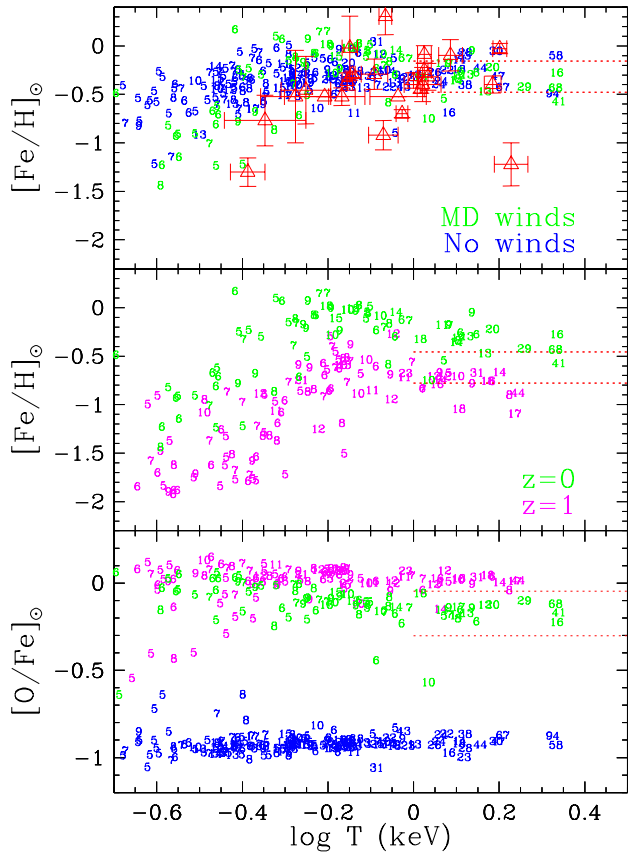}}
\vskip -0.5in
\caption{{\it Top:} [Fe/H]$_\odot$ ($L_X$-weighted)
for galaxy groups drawn from our
momentum-driven simulation (green) and no-wind simulation (blue) at $z=0$.  
Small numbers indicate the number of resolved galaxies in the group.  
Data points (red) from \citet{hel00} are shown for groups.
The observed cluster iron abundance range of $0.33-0.7$~solar~\citep{pet03} 
is indicated as horizontal dotted lines down to $T_X=1$~keV.
{\it Middle:} [Fe/H]$_\odot$ in the momentum-driven wind simulation
at $z=0$ (green, reproduced from above panel) and $z=1$ (magenta).
The observed range at $z=1$ for comparison is schematically indicated 
by the dotted lines showing $0.1-0.25$~solar, i.e. half of the $z=0$ abundances.
{\it Bottom:} [O/Fe]$_\odot$ in the momentum-driven (green) and
no-wind (blue) simulations at $z=0$, and the momentum-driven run
at $z=1$ (magenta).  Range of [O/Fe] observations for clusters at 
$z\approx 0$ is indicated by the dotted lines~\citep{pet03}.
}
\label{fig:metal}
\end{figure}

\subsection{Iron abundance}

Figure~\ref{fig:metal} (top panel) shows the iron abundance, weighted
by X-ray luminosity, as a function of $L_X$-weighted temperature
$T_X$ for simulated groups.  The green points show the results for our
momentum-driven wind simulation, and the blue points show the no-wind run.
The number of resolved galaxies in each group is indicated by the small
number on the plot (in this and subsequent figures).  The key result
is that the iron abundance hovers around the observed value between
one-third and two-thirds solar for $T_x\approx 0.5-2$~keV systems.
The range of observations for clusters from 1.5--10~keV ([Fe/H]$_\odot=0.33-0.7$) from
\citet{pet03} is shown as the dotted lines.  Individual iron abundances
for galaxy groups from \citet{hel00} are shown as the red points;
observations by \citet{davis99} show a similar trend.  There is a mild
trend for lower-mass systems to have higher metallicities down to 0.5~keV,
then a significant drop to smaller systems.  Qualitatively, this is in
agreement with observations, though data on the poorest systems is still
highly uncertain.

Matching the observed enrichment level of intragroup gas is a
significant success for this outflow model, and indicates that
our chemical enrichment model is working together with stellar mass
growth to properly enrich X-ray emitting gas.  \citet{tor07} achieved
similar success for clusters, with a similarly sophisticated chemical
enrichment model implemented in \gad.  Hence a metallicity of one-third
to one-half solar in intracluster and intragroup gas is achievable using
standard yields and initial mass functions, when the cosmic production
and distribution of metals is properly incorporated into hierarchical
structure formation models.  This suggests that non-standard metal yields
or initial mass functions~\citep[IMFs;][]{por04} are not required to match
observed abundance levels.  In fact, \citet{tor07} and \citet{fab08}
found that standard yields and a Salpeter IMF provides a better fit to
cluster abundances and its evolution than a highly top-heavy IMF.

Perhaps more surprising is that no wind simulation also roughly reproduces
the observed iron abundance.  One might surmise that this is because the
iron content in intragroup gas is dominated by delayed Type~Ia supernovae,
thereby tracking the stellar mass distribution, and gaseous outflows play
a subdominant role.  To some extent that is true, but it turns out that
the apparent insensitivity of iron enrichment to outflows is something of
a coincidence.  In reality, the no-wind simulation only matches the iron
abundance because it overproduces the amount of stars in groups, thereby
compensating for the lack of metal dissemination into intragroup gas.
We will return to this point when we examine the baryon fractions in
\S\ref{sec:baryfrac}.

The evolution of the iron abundance has now been measured for massive
clusters out to $z\sim 1$.  \citet{bal07} and \citet{mau08} found an
increase of roughly a factor of two from $z\approx 1\rightarrow 0$.
The middle panel of Figure~\ref{fig:metal} shows the iron abundance in
our momentum-driven case at $z=0$ (green points) and $z=1$ (magenta).
The dotted lines show the \citet{pet03} range at $z=0$ lowered by
a factor of two, for a rough comparison to the $z=1$ predictions.
Our simulation predicts iron abundance growth that is roughly consistent
with that seen in clusters, though the growth rate seems somewhat higher
in the models than in the data.  Of course we are comparing simulated
groups with observed rich clusters at $z=1$, so the discrepancy may
be due to differential metal evolution between groups than clusters.
But the predicted metal increase is independent of system size, at least
in the range we can probe with these simulations, so this explanation is
disfavored.  Another possibility is that our central group galaxies have
too much star formation at $z<1$, since we do not include any feedback
mechanism for truncation of star formation.  These minor discrepancies
aside, the predicted metallicity evolution is generally in agreement
with observations.

The outflow simulation predicts a significant drop in metallicity below
0.5~keV or so.  This is likely a reflection of the mass-metallicity
relation for galaxies setting in~\citep[e.g.][]{fin08}.  Poor groups of
this size typically house $\sim L_*$ galaxies (such as our Local Group),
and below $L_*$ the metallicity of a galaxy begins to drop with stellar
mass~\citep{tre04}.  Since the mass-metallicity relation is governed by
outflows, one would expect the no-wind simulation to show no such drop,
and Figure~\ref{fig:metal} confirms this to be the case.  Hence although
the star formation efficiency and stellar fractions are higher in
smaller groups (as we will see in \S\ref{sec:baryfrac}), outflows are
able to compensate for this by driving more metals out of group potentials.
\citet{davis99} and \citet{hel00} found that the poorest groups in their
sample at $T_X\la 1$~keV have lower metallicities (see data points at
$T<1$~keV in Figure~\ref{fig:metal}), but the samples are small and the
uncertainties large.  Our simulation also predicts an increase in the
spread in metallicities at low-$T_X$, owing to the stochastic nature
of outflows; this may help explain the large range of metallicities
seen for the poorest groups.  As observations improve in this regime,
the abundances of the smallest X-ray groups should provide a critical
test of outflow models.

\subsection{Oxygen abundance}

The oxygen to iron ratio is often employed as a diagnostic of the
contribution from Type~II versus Type~Ia supernovae, since oxygen is
produced primarily in the former while iron primarily in the latter (at
least in groups and clusters).  The bottom panel of Figure~\ref{fig:metal}
shows [O/Fe] for the momentum-driven wind simulation at $z=0$ (green
points) and $z=1$ (magenta).  These metallicities are X-ray emission
weighted, so they trace the enrichment level of hot gas only.  Since
oxygen enrichment occurs predominantly at earlier epochs coincident with
star formation, while Type~Ia supernovae enrich in iron over longer cosmic
timescales, [O/Fe] drops mildly with redshift.  Our wind simulations
predicts a $\sim 30$\% drop from $z=1\rightarrow 0$, independent of
system size.  Future observations of ICM alpha element abundances in
high-$z$ clusters can test this prediction.

Comparing [O/Fe] between our wind (green points) and no-wind (blue)
simulations, it is evident that the [O/Fe] ratio provides a strong
constraint on outflow properties.  Without outflows, oxygen is not
disseminated effectively into diffuse gas, and instead remain trapped
in stars.  In our no-wind run, this results in [O/Fe]$_\odot\approx -1$
at all group masses.  Observations of clusters from $1.5-10$~keV by
\citet{pet03} indicate an oxygen to iron ratio of $70\pm20$\% compared
to solar, with no trend versus system size; this range is indicated
by the dotted lines in the bottom panel of Figure~\ref{fig:metal}.
Our momentum-driven wind run places much more oxygen into X-ray emitting
gas, and no trend with system size, in remarkably good agreement with
observations.  It is notable that a different wind model explored by
\citet{tor07}, namely the ``constant wind" model of \citet{spr03b}, was
not able to achieve as high [O/Fe] ratios as observed.  Hence, [O/Fe]
represents a sensitive test of outflows, which our momentum-driven wind
simulation nicely passes.

Contrary to the usual assumption, we find that the [O/Fe] ratio in X-ray
emitting gas is not a good indicator of the relative numbers of Type~II
and Type~Ia supernovae that have exploded within the group potential.
Rather, [O/Fe] (and presumably other alpha-to-iron ratios) reflect
more the distributions mechanisms of the various elements.  This is
seen by comparing the bottom panels of Figure~\ref{fig:metal} and
Figure~\ref{fig:baryfrac}.  The latter shows the global mass-weighted
[O/Fe] ratio in our simulated groups, including both stars and cold
gas.  In both the no-wind and momentum-driven wind simulations, the
mass-weighted [O/Fe] is around solar, roughly independent of group size.
This is in stark contrast to typical values of 0.7~solar in the outflow
run, and 0.1~solar in the no-wind case.  This arises because oxygen is
predominantly distributed into hot gas via winds from galaxies, while iron
is predominantly distributed from old stars.  When no winds are present,
oxygen remains trapped within galaxies, eventually being locked into
stars, while iron comes from older stars that can dynamically diffuse
out of galaxies directly into hot gas.  As a result, X-ray weighted
[O/Fe] is a sensitive tracer of metal distribution mechanisms, but is
not effective as a stellar chronometer.

At face value, our results further imply that tidal and ram pressure
stripping of metals out of galaxies falling into deep potential wells is
not sufficient to pollute intragroup gas as observed.  Such dynamical
processes are in principle implicitly included in the no-wind run, yet
the no-wind run fails to enrich intragroup gas sufficiently with oxygen.
This result apparently contradicts the semi-analytic+N-body simulation
results of \citet{kap07}, whose models predict that ram pressure stripping
dominates hot gas enrichment for present-day clusters.  This may indicate
that the simplified modeling used in that work is not capturing the full
hydrodynamics.  Alternatively, it may be that ram-pressure stripping is
numerically suppressed in our simulations.  For instance, \citet{age07}
demonstrated that \gad\ tends to oversuppress stripping relative to
AMR codes, as SPH has trouble modeling hydrodynamical instabilities.
Hence while our model of dynamical metal ejection is successful, further
testing is required to determine whether gas stripping processes can
also be effective.

\subsection{Baryon fractions}\label{sec:baryfrac}

\begin{figure}
\vskip -0.3in
\setlength{\epsfxsize}{0.6\textwidth}
\centerline{\epsfbox{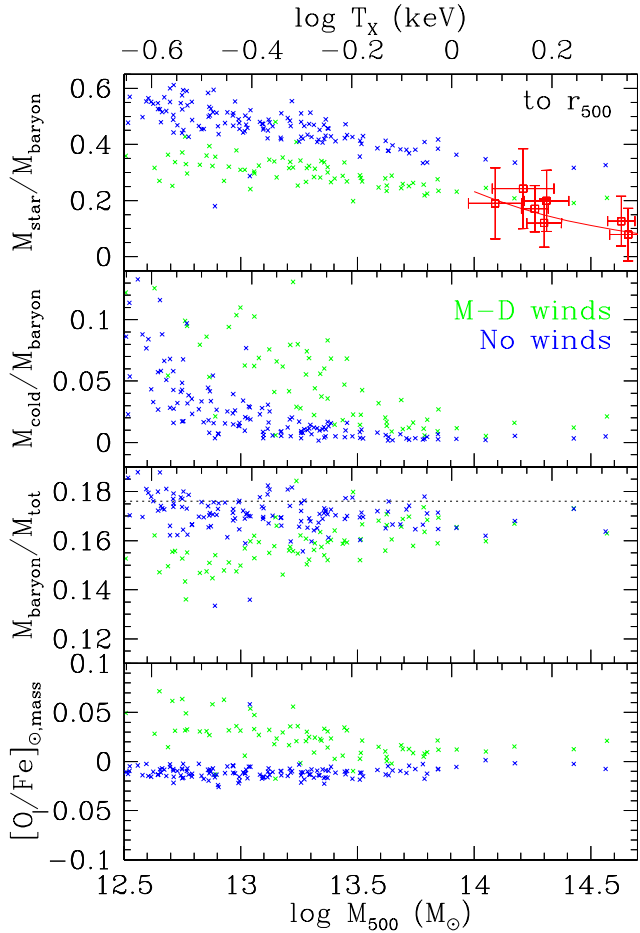}}
\vskip -0.5in
\caption{Baryon fractions is various phases, out to $r_{500}$.
Green points are for our favored momentum-driven wind simulation,
while blue points are for the no-wind run.  
{\it Top:} Stellar baryon fraction, including all stars in the groups
regardless of whether they are within galaxies or not.  Red points and 
red line show data and fit to observations of stellar baryon fractions
from \citet{gon07}; their data for groups $<10^{14}M_\odot$ are not shown 
because they lie beyond the range where the $M_{500}-r_{500}$ relation 
has been calibrated, so are highly uncertain (A. Gonzalez, priv. comm.).
Top axis shows approximate X-ray temperature based on a fit to 
$M_{500}-T_X$ from our simulations.
{\it Second:} Cold baryon fraction, where cold gas is all gas below
$10^6$K.
{\it Third:} Baryon fraction with respect to total mass.  Dotted line
shows the ratio $\Omega_b/\Omega_m=0.176$ assumed in our simulation.
{\it Bottom:} Log of oxygen to iron mass to $r_{500}$, including both
gas and stars, relative to solar ratio.  This can be compared to the
$L_X$-weighted [O/Fe] in Figure~\ref{fig:metal}.
}
\label{fig:baryfrac}
\end{figure}

Figure~\ref{fig:baryfrac} shows the baryon fractions as a function of
halo mass out to $r_{500}$ in our simulations with winds (green points)
and without (blue).  The top panel shows the fraction of baryons in stars
(we do not distinguish stars in galaxies from intracluster/intragroup
stars).  It is evident that galactic outflows are effective at suppressing
star formation on group mass scales.  Observational estimates of the
stellar fraction on group scales are challenging, primarily owing to the
difficulty in constraining the dominant baryonic component in hot gas, but
data from \citet{gon07}, which carefully accounts for intracluster stars,
are broadly reproduced in the momementum-driven wind run.  As seen in the
second panel, there is virtually no cold gas ($T<10^6$K) at least in the
range of observed X-ray groups, so baryons not in stars are essentially
all in an X-ray emitting plasma.  The \citet{gon07} stellar fraction data
are consistent with data from \citet{lin04} for $M_{500}>10^{14}M_\odot$
when corrected for ICL light contributions~\citep{bal08}.  Below this
mass, systematic uncertainties become large.

The most massive groups still have 20\% of their baryons in stars, which
is too high compared to data, and a simple extrapolation of the simulated
group trends to higher masses would clearly violate observations of
stellar fractions in rich clusters (typically $\ll 10\%$).  Another way
to say this is that the observed $f_*-M_{500}$ relation (where $f_*$
is the stellar fraction) is steeper than our simulations predict.
Specifically, our simulations yield $d\log f_*/d\log M_{500}=0.18$
around $10^{14}M_\odot$, which is shallower than both \citet{gon07}
and \citet{lin04} data, though somewhat steeper than the semi-analytic
model predictions of \citet{bow06}~\citep[as calculated by][]{bal08}.
However, this does not necessarily indicate a significant constraint
on cold dark matter models~\citep[as argued by][]{bal08}.  Rather,
it probably owes to a lack of suppression of star formation in massive
systems in simulations, from AGN feedback or whatever the heating source
is that prevents cooling flows.

The suppression of star formation on group scales is not as great as the
global suppression:  The simulation without winds at $z=0$ contains 17\%
of its total baryons in stars, while the wind run has 7\%.  The latter
value is in good agreement with observational estimates~\citep{bel03,bal08}.
Hence outflows are able to suppress star formation both globally and on
group scales in general agreement with observations.

The third panel of Figure~\ref{fig:baryfrac} shows the mass fraction of
total baryons against halo mass.  In general, the total baryon fraction is
slightly smaller than the cosmic baryon fraction (indicated as the dotted
line), owing to increased pressure support that puffs out the hot gas;
this is qualitatively consistent with observations~\citep[e.g.][]{gon07}.
Outflows are further able to remove a small but noticeable amount of
baryons from halos with masses $\la 10^{14}M_\odot$.

Returning to the iron abundance with and without outflows, it is now
clear that although simulations without outflows reproduce the observed
intragroup metallicities, they do so by forming too many stellar baryons.
Outflows therefore serve the dual purpose of suppressing star formation,
while transferring sufficient metals from galaxies into the hot gas phase.
Furthermore, outflows significantly increase the alpha element ratio in
intragroup gas to be in better agreement with observations.  Hence our
simulations suggest that outflows are likely to be necessary in order
to enrich intragroup gas to observed levels.

\section{Pre-heating}\label{sec:entropy}

We now turn to studying another aspect of outflows, namely the energy they
add to intragroup gas.  Entropy levels in intragroup gas are known to be
elevated relative to self-similar extrapolations from clusters, but the
physical process responsible for pre-heating is unclear. \citet{voi01}
argued analytically that supernova heating together with radiative
cooling could yield the observed entropy distribution in groups.
Incorporating such feedback processes into simulations has only recently
been attempted; for instance, \citet{bor05} examined and dismissed a
simple model of galactic outflows as the dominant source of pre-heating.
\citet{mua06} showed that an ad hoc model of energy input proportional
to star formation is able to reproduce observed entropy levels,
but preliminary comparisons to entropy evolution favored an early
pre-heating scenario.  None of these works employed an outflow model that
is constrained to match external observations.  Here we examine whether
our momentum-driven wind model, in addition to enriching intragroup gas,
is concurrently able to add significantly to its entropy budget.

\subsection{Entropy}\label{sec:coreentropy}

\begin{figure}
\vskip -0.5in
\setlength{\epsfxsize}{0.7\textwidth}
\centerline{\epsfbox{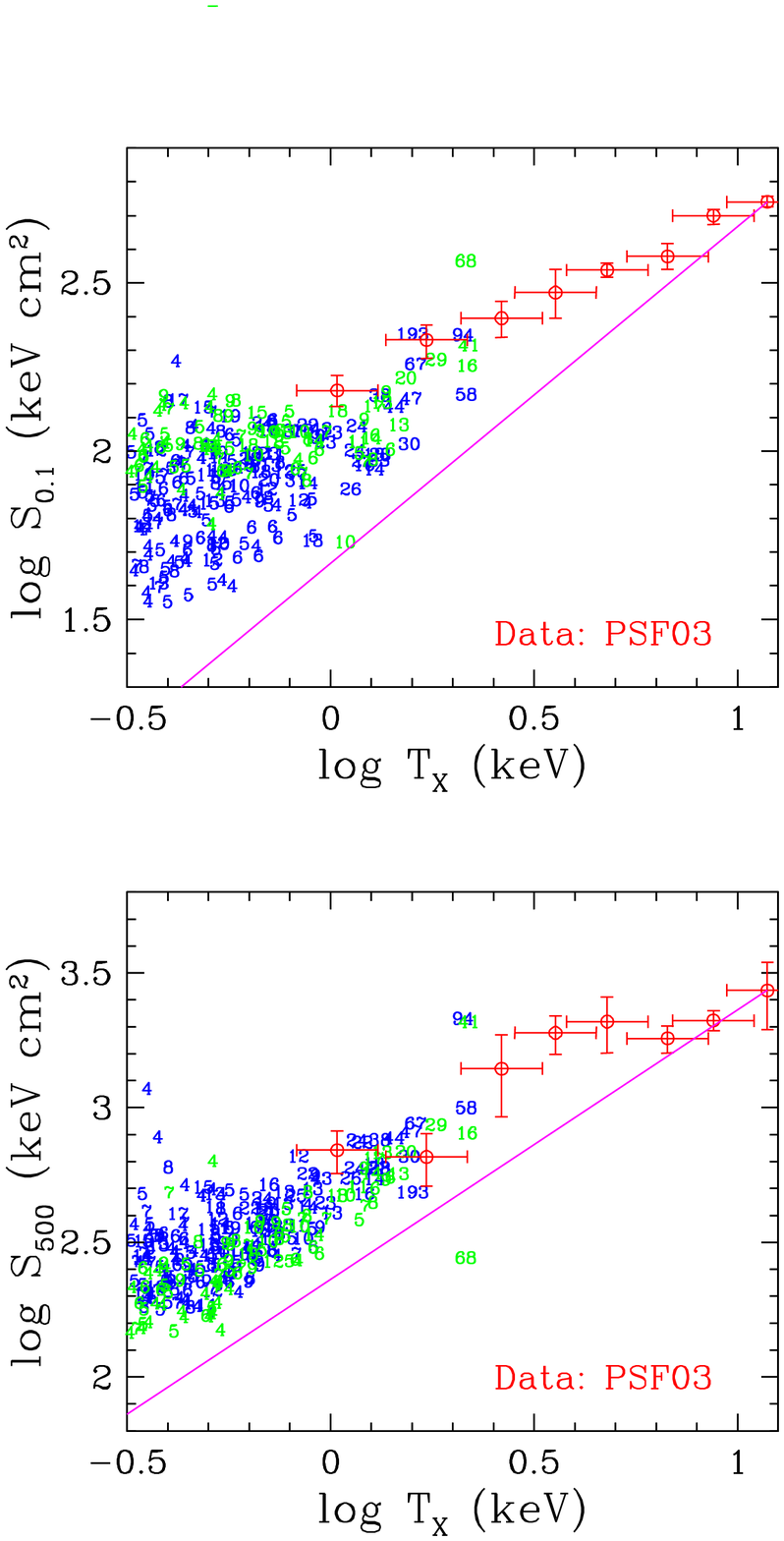}}
\vskip -0.5in
\caption{ Entropy, defined as $T_X/n_e^{2/3}$, as a function of
X-ray temperature for galaxy groups in our simulation with outflows
(green) and without (blue).  Binned data from PSF03 is shown
in red.  Expected self-similar scaling, normalized to the most massive data
points, is shown as the magenta line.
{\it Top:} Entropy at 10\% of the virial radius.
{\it Bottom:} Entropy at $r_{500}$, roughly two-thirds the virial radius.
}
\label{fig:Stemp}
\end{figure}

Figure~\ref{fig:Stemp} (top panel) shows entropy $S_{0.1}$ at $0.1R_{\rm
vir}$ as a function of group X-ray temperature, for our momentum-driven
wind (green) and no-wind (blue) runs.  Entropy is defined here as
$T_X/n_e^{2/3}$, where $T_X$ is the system X-ray luminosity-weighted
temperature and $n_e$ is the electron density at a given radius.
Data points (binned) from PSF03 are shown, and the expected self-similar
scaling normalized to the most massive clusters is indicated by
the magenta line.  In self-similar scaling, $n_e(0.1R_{\rm vir})$ is
independent of size since groups are scaled-down versions of clusters,
hence $S\propto T_X$.  As is well-known, $S_{0.1}$ is elevated at group
scales relative to a self-similar extrapolation from massive systems.
While an entropy floor at $\sim 100$~keV~cm$^2$~\citep{pon99} is not
seen in the models, more recent observations indicate no floor but
merely a more gradual decline than expected.  PSF03 finds a best fit of
approximately $S_{0.1}\propto T_X^{2/3}$.

Elevated entropies are obtained in both wind and no-wind runs, since
radiative cooling is effective at preferentially removing low-entropy
intragroup gas over a Hubble time \citep[e.g.][]{bry00,voi01}.
The low-entropy gas then ends up forming stars.  While the entropy
trend produced by radiative cooling qualitatively agrees with data,
as shown in \citet{dav02}, there are two major difficulties with the
radiative cooling solution alone.  The first is that the amount of
stars produced by the low-entropy gas exceeds observations, as shown in
Figure~\ref{fig:baryfrac} and found previously~\citep[e.g.][]{mua02}.
The second is that, as seen in Figure~\ref{fig:Stemp}, the entropy
increase from radiative cooling alone is not quite sufficient to match
observations.  Hence an alternate source of entropy is still required,
preferably one that simultaneously reduces the amount of star formation
in groups.

Our momentum-driven outflow model appears to alleviate both problems.
We have seen that it reduces the stellar baryon fraction to be in better
agreement with observations on group scales.  Figure~\ref{fig:Stemp}
also shows that it raises the entropy level over the no-wind case,
particularly for systems with $T_X\la 1$~keV, to be in slighly
better agreement with the low-$T_X$ end of available observations.
An extrapolation of the observed trend ($S\propto T_X^{2/3}$) to lower
$T_X$ is also well-matched to the outflow run predictions, which can
be fit by $S_{0.1}\propto T_X^{0.6}$ over the temperature range probed.
Hence outflows together with radiative cooling appear to provide a viable
solution to the entropy crisis in galaxy groups.

The fact that outflows {\it increase} core entropy is not trivial,
since they also suppress cooling, and hence suppress the removal of
low-entropy gas.  The entropy addition from outflows therefore appears to
be (more than) compensating for the reduced stellar fraction.  That said,
our most massive systems show twice the stellar fraction as observed
(see Figure~\ref{fig:baryfrac}), and so some additional mechanism such
as AGN feedback is likely required to further suppress star formation at
$T_X\ga 1$~keV.  It is not clear whether such a mechanism would continue
to compensate for further reductions in stellar fractions in the way that
outflows do.  Looking carefully at Figure~\ref{fig:Stemp}, it does appear
that the core entropy is still slightly low for $T_X\ga 1$~keV systems,
though it is within uncertainties.  Hence it could be that radiative
cooling, outflows, and some other feedback mechanism all work together
to establish the entropy level in larger groups and clusters.  A fully
self-consistent model incorporating all these processes is required to
assess this.

The entropy in group outskirts is also seen to be elevated with respect
to a self-similar extrapolation from clusters.  The bottom panel of
Figure~\ref{fig:Stemp} shows the entropy at $r_{500}$, roughly two-thirds
the virial radius.  Observations (PSF03, shown as the red points)
are more uncertain here, particularly for groups, but our simulations
generally fall in the observed range.  $S_{500}$ is mostly insensitive to
outflows; it is apparently established by radiative cooling.  If anything,
the wind run shows marginally lower entropies, probably owing to reduced
removal of low-entropy baryons.  The $S_{500}-T_X$ relation shows a slope
just below unity, unlike $S_{0.1}$ and more in line with expectations
from self-similarity.  If this would continue to the largest clusters,
it would exceed the observed $S_{500}$; this will require constrained
realization simulations of large clusters to test.

\subsection{Outflow energy deposition}\label{sec:energy}

We now quantify pre-heating due to outflow energy injection in
simulations.  During our runs we record the mass and velocity (and hence
energy) of particles kicked into an outflow.  Ideally, one would like
to track where and when this energy gets deposited into surrounding gas;
unfortunately, this is not straightforward.  So instead, we make a crude
estimate of energy input into intragroup gas by assuming that all the
outflow particles lying within a group's hot gas by $z=0$ have deposited their
entire outflow energy into that gas.  Note that this estimate
may either be an underestimate of the actual energy added if outflow
particles lose energy to intragroup gas but end up outside the group,
or it may be an overestimate if groups accrete outflow particles that
already have deposited most of their energy outside the group's halo.
While this estimate is not exact, it is still instructive.

\begin{figure}
\vskip -0.5in
\setlength{\epsfxsize}{0.7\textwidth}
\centerline{\epsfbox{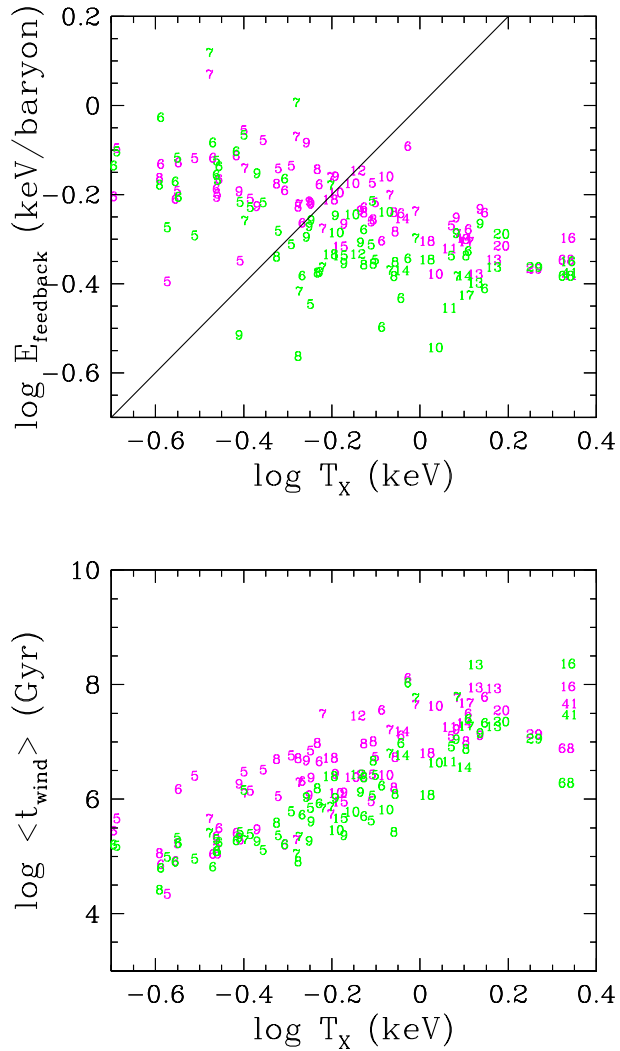}}
\vskip -0.5in
\caption{{\it Top:} Estimated energy per baryon deposited into intragroup gas 
from galactic outflows (green points), as a function of system temperature.  
Solid line is the line of equality between wind energy and system thermal 
energy.  Magenta points show supernova energy per hot baryon from all the 
stars formed within the groups.
{\it Bottom:} Age by which half the energy was deposited (green points),
taken to be the median ejection time of all wind particles in the hot
intragroup gas.  Magenta points show median age of stars in group.
}
\label{fig:ewind}
\end{figure}

Figure~\ref{fig:ewind} shows the wind energy per hot baryon deposited as
a function of $T_X$ (top panel), and the median age when the energy was
deposited (bottom panel), as the green points.  The energy is the total
outflow energy of all wind particles residing within the hot intragroup
gas, divided by the number of baryons in that gas (note that this is
not all the baryons in the group).  The age is estimated by taking the
median of the ejection times of all wind particles within intragroup gas;
again, this must be considered a relatively crude estimate of the energy
injection time.

The energy deposition per baryon shows a mild decreasing trend with system
size.  This is at first glance surprising, since in the momentum-driven
outflow model the wind energy output per unit star formation scales as
$\eta\vw^2\propto v_{\rm circ}$ (equation~\ref{eqn:vwind}).   Hence one
expects larger systems to have larger energy input.  But the baryonic
budget serves to counter this trend:  Larger groups have a lower
stellar fraction and more hot gas over which to spread the energy
(Figure~\ref{fig:baryfrac}).  The net result is that hot baryons in
larger systems are heated slightly less by winds.

The total energy injection into groups from outflows is comparable to
the total supernova energy from stars formed in the group (magenta
points in Figure~\ref{fig:ewind}).  The SN energy is computed by
taking the total stellar mass in each group and multiplying by $4\times
10^{48}$~ergs$/M_\odot$ for a standard IMF~(OD08).  As argued in OD08,
such high levels of energy input appear to be necessary to provide
the cosmic distribution of metals from galactic outflows, and owing to
expected radiative losses this level of energy may not be obtainable
from supernovae alone \citep[barring IMF variations; e.g.][]{dav08}.
While we cannot claim to fully understand or model the outflow driving
mechanism, we note that such energetic outflows seem to be required to
match a wide range of observations, not just intragroup gas entropy.

The median energy injection time (green points in bottom panel of
Figure~\ref{fig:ewind}) is typically fairly recent, with the smallest
systems having a median pre-heating age of 5~Gyr ($z\approx 0.5$),
while our largest systems at around 2~keV have a typical pre-heating
age of 8~Gyr ($z\approx 1$).  This can be compared to the median age
of star particles in the group (magenta points), which are comparable
although slightly older, particularly in the smaller systems.  In general
it appears that stars in group galaxies are responsible for enriching
and pre-heating intragroup gas, as expected.  In the future we plan to
examine the redshift evolution of hot gas to determine more precisely
where and when outflow energy and metals are deposited.

\subsection{Scaling relations}

Figure~\ref{fig:ewind} shows that the energy deposition from winds
typically exceeds the system's thermal energy for systems below about
0.5~keV (note the solid line showing equality between $T_X$ and wind
energy input).  Hence outflows should produce a lowered luminosity for
systems near this temperature and below.  Here we investigate whether
this provides an explanation for the apparent break in luminosity
scaling relations near $\sim 1$~keV, below which the $L_X-T_X$ relation
is observed to be significantly steeper~\citep[e.g.][]{xue00}.

\begin{figure}
\vskip -0.5in
\setlength{\epsfxsize}{0.75\textwidth}
\centerline{\epsfbox{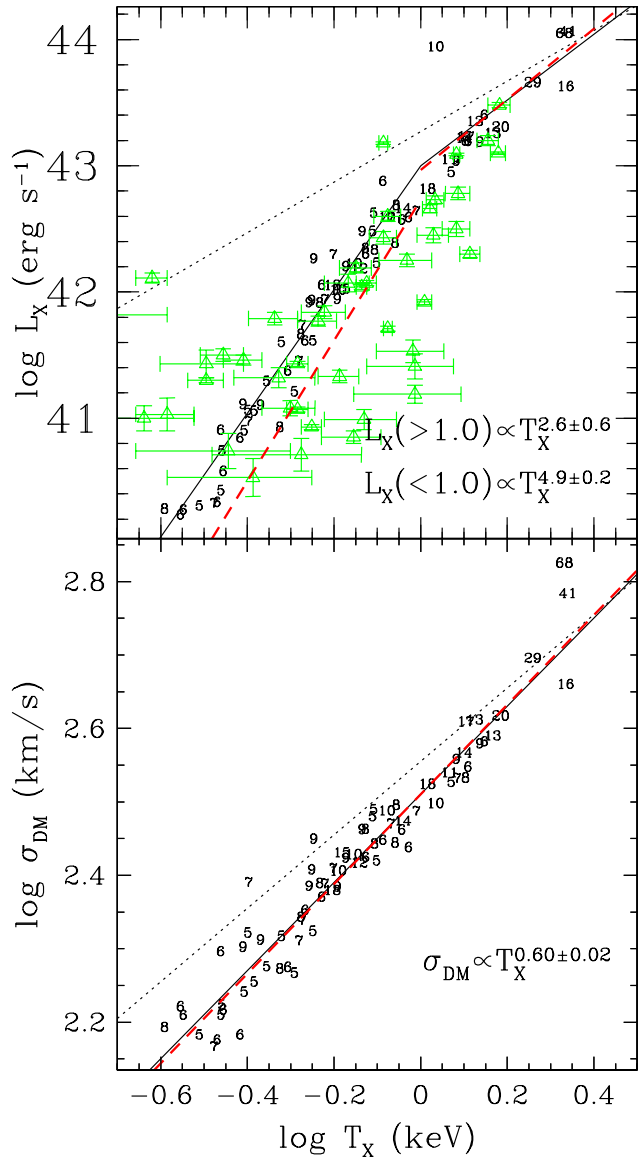}}
\vskip -0.4in
\caption{$L_X-T_X$ (top panel) and $\sigma-T_X$ (bottom) relations for
simulated groups.  Power-law fits are indicated in the lower right,
and the fits are shown as the solid lines.
Dotted lines show self-similar scaling normalized to the largest systems.
Fits to observations of groups and clusters compiled by \citet{xue00}
are shown as the red dashed lines.  Green triangles show individual group
observations from \citet{osm04}.  
}
\label{fig:txscale}
\end{figure}

Figure~\ref{fig:txscale} shows the $L_X-T_X$ (top) and $\sigma-T_X$
(bottom) relations for our simulated groups from our momentum-driven wind
run.  The $L_X-T_X$ relation shows a clear break around 1~keV, with the
power-law slope above and below this break being $2.6\pm0.6$ and $4.9\pm
0.2$, respectively.  These slopes agree well with observations: The
red dashed line shows a fit to observations of groups and clusters from
\citet{xue00}; a fit to $L_X-T_X$ for clusters by \citet[not shown]{whi97}
is quite similar.  Green points show data from a ROSAT compilation
by \citet{osm04}; we show all their groups that have measured $T_X$
values, including systems with a small radial extent in observable X-rays
(i.e. their ``H-sample") that they exclude from their fits, since we make
no such cuts in our simulations.  The $L_X$ values used are extrapolated
to $r_{500}$, whereas our simulations include all luminosity out to
the virial radius, but this should not be a large difference.

The \citet{osm04} $L_X-T-X$ data as well as the \citet{xue00} fit at
$T_X<1$~keV lie systematically below that for the simulated groups.
If real, this suggests that there may still be some missing energetic
input into low-mass systems.

The no-wind simulation (not shown) also displays a break in the $L_X-T_X$
relation, but it is not as prominent and occurs at a lower $T_X$.
Our no-wind case is in general agreement with simulations (also without
outflows) by \citet{dav02}, who argued that cooling alone can produce
a break in scaling relations, but this break occurs at $\sim 0.3$~keV.
Hence outflows serve to add enough entropy to move the $L_X$ break closer
to the observed location.

The $\sigma-T_X$ relation (bottom panel of Figure~\ref{fig:txscale}) shows
no break.  Here, $\sigma$ is calculated from the dark matter particles
in each simulated group; in \citet{dav02} it was shown that calculating
$\sigma$ from galaxies yields a similar values so long as there are a
sufficient number of galaxies in a group.  The simulated groups are best
fit by a power-law slope of $0.60\pm0.02$ (solid line), and the fit is
remarkably similar to that based on compiled observations of groups and
clusters by \citet[red dashed line]{xue00}.  This is slightly different
from the self-similar scaling slope of 0.5, and shows that groups at
a given mass have elevated temperatures.  The no-wind simulation (not
shown) shows a similar slope~\citep[see][]{dav02}, indicating that the
elevated temperatures occur primarily because radiative cooling has
preferentially removed the low-temperature gas.

The scatter in the observational points is considerably larger than in
the simulated case.  As an example, green points (with errors) on the
$L_X-T_X$ plot show individual measurements from \citet{osm04}.  The large
scatter may be due partly to observational systematic uncertainties,
but this seems unlikely to provide the bulk of the explanation, since
clusters show similar scatter that can be directly linked to their
well-measured density and temperature profiles~\citep[e.g.][]{mar98}.
It has been suggested that it may instead reflect intrinsic scatter in
the level of pre-heating~\citep{bal06}.  Since our simulations implicitly
produce variations in halo concentration and formation epoch, as well
as in wind energy deposition, these alone cannot provide the observed
scatter~\citep[as also argued by][]{bal06}.  We leave for the future a
careful consideration of observational causes of scatter via making mock
X-ray images and analyzing them alongside real X-ray data.  This will
also be useful for evaluating the more general bias that observed groups
tend to be X-ray bright, but these may not be representative of groups
as a whole.  The final solution, however, may require some intermittent
additional source of heating such as AGN feedback.

In summary, outflows add an important contribution to the entropy
budget of poor groups, whose energy input becomes comparable to the
system thermal energy at $T_X\sim 1$~keV.  Radiative cooling is primarily
responsible for breaking self-similarity above this scale, while outflows
exaggerate deviations from self-similarity below this scale.

\section{Profiles}\label{sec:profiles}

\begin{figure}
\vskip -0.5in
\setlength{\epsfxsize}{0.65\textwidth}
\centerline{\epsfbox{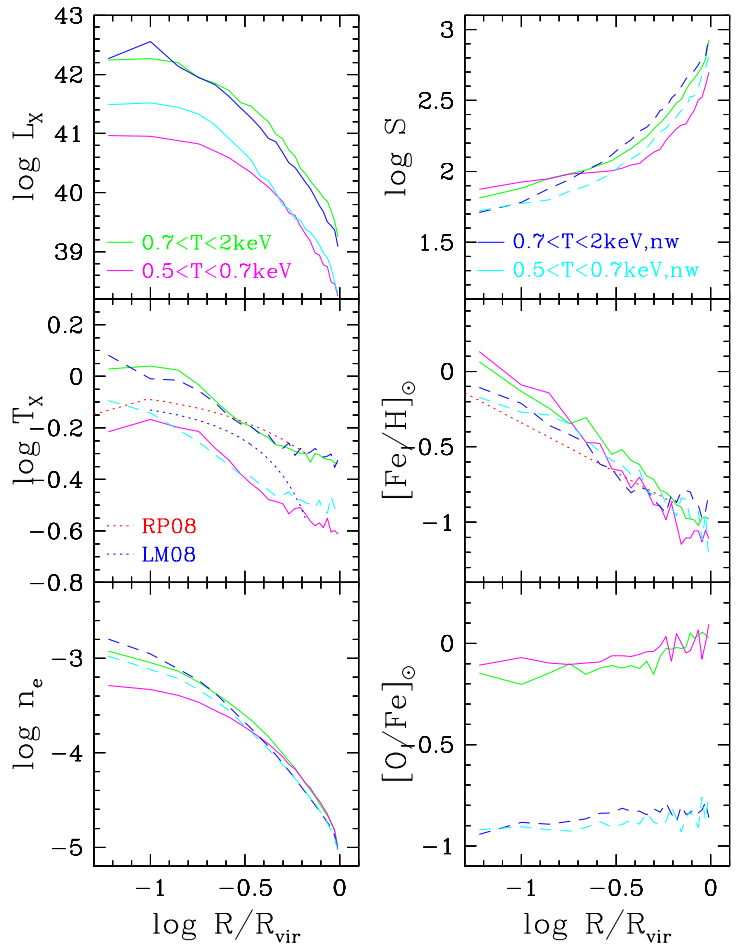}}
\vskip -0.5in
\caption{Radial profiles of physical quantities in our momentum-driven
wind simulation.  Profiles shown are medians over all galaxy groups 
within two bins: $0.7<T_X<2$~keV (green) and $0.5<T_X<0.7$~keV (magenta);
each bin contains $\approx40$ groups.  
Blue and cyan curves show median profiles of groups in those two
temperature ranges from the no-wind simulation.
Note that there is significant scatter in individual profile shapes,
which is not shown.  Dotted lines in $T_X$ and [Fe/H] panels shows
observations; red are from \citet{ras07}, blue are from \citet{lec08}.
{\it Top:} X-ray luminosity (erg/s);
{\it Second:} X-ray luminosity-weighted temperature (keV);
{\it Third:} Free electron density (cm$^{-3}$);
{\it Fourth:} Entropy (keV~cm$^2$);
{\it Fifth:} Luminosity-weighted iron abundance; 
{\it Bottom:} Oxygen to iron ratio.
}
\label{fig:xprof}
\end{figure}

With increasingly sensitive observations from {\it Chandra} and {\it XMM},
it is now possible to construct radial profiles of group properties.
Such profiles provide an added challenge to models, one that simulations
have traditionally had a difficult time matching, since they contain
more detailed information about metal and energy injection processes.
In this section we examine projected radial profiles for physical
quantities from our momentum-driven wind simulation.

Figure~\ref{fig:xprof} shows the luminosity, temperature, electron
density, entropy, iron abundance, and oxygen to iron ratio profiles as a
function of radius, from $\approx 0.06-1 R_{\rm vir}$.  The inner limit
corresponds to $\approx 0.1R_{500}$, which is typically few tens of
kpc, and approaches the effective resolution limit of our simulations.
As seen in the top panel, the luminosity profile shows a flat central
core~\citep[similar to observed non-cool core systems; see][]{dav02},
so the poorly resolved central region does not make a significant
contribution to groups' X-ray emission-weighted properties.

We subdivide the groups into two equal-number bins exceeding 0.5~keV,
which results in about 40 groups each from $0.5-0.7$~keV and $0.7-2$~keV.
We then compute the median value among the 40 groups for each physical
quantity within each of the 20 radial bins.  This yields the median
profiles shown within these two temperature ranges.  We show such profiles
for our momentum-driven wind run (green and magenta) and no-wind run
(blue and cyan).

The simulated temperature profiles show a drop by a factor of $\sim
2.5$ from $0.1\rightarrow 1R_{\rm vir}$.  The no-wind profiles are
well-described by the fit proposed by \citet{lok02}, namely $T_X\propto
(1+0.75 r/r_{500})^{-1.6}$.  With outflows, the profiles are altered
slightly in the inner parts, such that they show a flattening and
perhaps even a turnover within $\sim 0.2R_{\rm vir}$.  Qualitatively,
this is in better agreement with X-ray group profiles from \citet{ras07},
who find a turnover in the temperature profile at $\sim 0.1 R_{\rm 500}$.

Fits to observed temperature profiles ($T/<T>$)are shown as the red and
blue dotted lines in Figure~\ref{fig:xprof}, arbitrarily scaled to lie
in the range of our simulated groups.  \citet[red dashed line]{ras07}
measured profiles for $0.3-2.1$~keV groups, so this is most directly
comparable to our simulated systems.  We show the \citet{lok02}
form of the fit to their data, specifically $T_X\propto (1+0.75
r/r_{500})^{-1.45}$, which is somewhat shallower than the simulated
profiles.  We also show a fit to cluster profiles by \citet[blue dashed
line]{lec08}.  Their analysis is notable because of the careful accounting
of various systematic effects such as background contamination and biases
from low surface brightness regions~\citep{lec07}, but other groups'
results are similar~\citep{deg02,vik05}.  Although the observed clusters
are not as directly comparable to our simulated groups, they may be more
accurate particularly in the outer regions, and since the profile shapes
seem to be mostly independent of system size, they may still be relevant.

The simulated $T_X$ profiles appear mildly steeper than observations
at $r>0.1R_{vir}$, as has historically been seen in such comparisons,
\citep[e.g.][]{mar98}.  There may be some additional source of
thermalization such as cosmic rays or conduction that serves to flatten
the real temperature profiles.  So long as such a process only moves heat
and not mass or metals around, then the remaining physical profiles should
not be significantly affected by fixing this discrepancy.  At smaller
radii, we do not have the resolution to reliably probe the temperature
profile too far in, but the observed profiles show a drop towards the
center that is similar to but steeper than seen in simulations (at least
for the larger groups).  This may result from an excess of star formation
in the simulated central group galaxies are over-heating the core.
Overall, the agreement with temperature profiles are reasonably
good but not perfect.

The observed iron abundance profiles are in reasonable agreement with
observations.  There is a hint that the simulated abundance profiles
are steeper than observed, but in this case the uncertainties are large
enough (particularly in the outer regions) that the differences are
probably not significant.  Our simulations may also not be resolving
the outer iron abundance profile ($r\ga 0.3R_{\rm vir}$) at our numerical
resolution~\citep{tor07}.  Outflows do not greatly impact the iron
abundance profile, because the stellar distribution is not altered
significantly by outflows even though they suppress star formation
overall.

The hot gas density profile shows a clear impact from outflows at
$T_X\la 1$~keV.  The $n_e$ profile is significantly shallower in the
central region for lower-mass groups in the outflow run.  In contrast,
the no-wind run shows little difference between the two mass bins,
although the increased efficiency of low-entropy gas removal does
cause the smaller systems to have slightly flatter inner profiles.
This demonstrates that outflows are important for further breaking
self-similarity in the core regions of poor groups as observed.

The entropy profiles are also affected by outflows.  The profiles from
the wind run are significantly shallower than those from the no-wind run.
Furthermore, less massive groups show a shallower profile relative to
more massive systems.  The outer profiles generally approach a slope of
1.1 expected for shock heating, but smaller systems do so further out.
PSF03 measured entropy profiles that look qualitatively similar, showing
a slope shallower than 1.1 out to $R_{500}$, particularly for smaller
systems.

Finally, the alpha enrichment profiles (here traced by [O/Fe]) show
a clear difference in amplitude between no-wind and wind runs.
This reflects the overall alpha enrichment difference seen in
Figure~\ref{fig:metal}.  More subtly, the no-wind runs show almost
no gradient in [O/Fe], while a mild gradient is seen in the wind run.
Still, the gradient is generally not as strong as observed in groups.
\citet{ras07} found that silicon to iron increased by a factor of 2.5 from
$0.1-1R_{500}$, whereas our predicted [O/Fe] only increases by 40\% over
a comparable range.  This may be owing to excess central star formation
in our simulated groups that raises the central oxygen abundance.

Overall, the X-ray profiles give a more detailed view of the physics
affected by outflow energy input as a function of mass.  Winds are
seen to clearly affect the gas density profile (and hence the entropy
profile), as well as the alpha element abundance profile.  Comparisons
with observations yield general agreement, though in detail significant
differences remain.  Hence these simulations do not completely explain all
observed group properties, even though global X-ray luminosity-weighted
properties are fairly well reproduced.  As detailed X-ray profile
observations of groups improve, matching them should yield insights
into other feedback or heat conduction processes operating within
intragroup gas.

\section{Conclusions}\label{sec:summary}

Using $\Lambda$CDM cosmological hydrodynamic simulations of galaxy
formation including galactic outflows and a sophisticated chemical
enrichment model, we have studied the distribution and evolution of
metals and entropy in galaxy groups.  The one-line summary is that our
simulation with outflow scalings as expected for momentum-driven winds
is broadly successful at reproducing both the enrichment and entropy
level seen in X-ray emitting intragroup gas.  In more detail, we find:

\begin{itemize}

\item Our model with outflows reproduces the observed iron abundance
of intragroup gas, while maintaining a stellar baryon fraction in broad
agreement with observed values.  The iron abundance is somewhat higher
than seen for rich clusters, but then drops for $T_X\la 0.5$~keV systems.

\item Predictions for [O/Fe] in X-ray emitting gas are strongly
affected by outflows.  Our momentum-driven wind model is successful at
reproducing the [O/Fe] ratio observed for clusters.  Without outflows
the oxygen abundance is too low, although it may be that our SPH-based
code is underestimating the effectiveness of gas (and metal) stripping.
Measurements of [O/Fe] in the poor group regime should provide stringent
constraints on outflow models.

\item The mass-weighted and X-ray luminosity weighted [O/Fe] ratio
within groups can be significantly different.  This suggests that [O/Fe]
in X-ray luminous gas is a poor indicator of the relative contributions
of Type~II and Type~Ia supernovae within the group as a whole, and is
instead governed primarily by the effectiveness of distribution mechanisms
for oxygen and iron.

\item Iron abundance evolution predicted for groups is in general
agreement with that seen for rich clusters, i.e. roughly a factor of
two increase from $z=1\rightarrow 0$.  Our simulations predict
that [O/Fe] decreases by $\sim 30\%$ over the same time span, 
though halting late-time star formation in central group galaxies
may increase such evolution.

\item The total baryon fraction in groups is predicted to be somewhat
below the cosmic mean baryon fraction, owing to pressure support from
hot gas.  In our outflow model this deficit increases in lower mass
systems owing to ejection of halo baryons, while without winds the baryon
fraction exceeds the cosmic value in the poorest groups.

\item The core ($0.1R_{\rm vir}$) entropy level of poor groups is found
to scale as $S_{0.1}\propto T_X^{0.6}$, in agreement with observations
by \citet{pon03}.  Much of the entropy increase over self-similar scaling
owes to radiative cooling alone, but outflows add significant entropy
to systems below 1~keV.

\item Entropy at $R_{500}$ is elevated from self-similar scalings,
in broad agreement with observations.  However, the scaling with
temperature within the poor group regime is close to self-similarity.
Outflows have essentially no impact here, as the elevation of $S_{500}$
owes purely to the removal of low-entropy gas via radiative cooling.

\item The amount of energy per baryon added to hot intragroup gas drops
with increasing system size, and exceeds the thermal energy below 0.5~keV.
As a result, the $L_X-T_X$ relation shows a clear break below 1~keV.
Both the $L_X-T_X$ and $\sigma-T_X$ relations predicted from our
momentum-driven wind run are in good agreement with observations.

\item Profiles of $T_X$ and [Fe/H] are qualitatively similar to observed,
though model profiles appear to be somewhat steeper than data.  The alpha
enhancement profile is also quite shallow.  Possible causes include
missing heat transport mechanisms and an excess of late-time star
formation in the central galaxy.

\item Outflow energy injection makes the density and entropy profiles
shallower at $T_X\la 1$~keV, particularly in the inner regions.

\end{itemize}

The simultaneous success of enrichment and entropy injection from our
momentum-driven wind outflows is a significant achievement, especially
considering that this outflow model was not tuned in any way to match
observations of intragroup gas.  While the iron abundance is not
very sensitive to outflows, the oxygen to iron ratio provides a key
diagnostic that is well reproduced by our outflow model.  Most of the
entropy increase actually owes to radiative cooling, but outflows serve
both to enrich intragroup gas as well as lower the stellar content to
bring it more into line with observations.

The implication of our models is that pre-heating in intragroup gas
arises from radiative cooling together with the same galactic outflows
required to, for instance, enrich the IGM at high redshifts or establish
the mass-metallicity relationship of galaxies.  Hence the source of
pre-heating may not be a mysterious widespread energy injection at
high redshifts, nor be related to AGN feedback at lower redshifts.
While something like AGN feedback is still required to truncate star
formation in the central galaxy, our results suggest that it need not be a
significant contributor to the overall entropy budget of intragroup gas.
This eases the tension between a feedback process that must operate on
dense central gas to prevent cooling flows (and therefore requires large
amounts of energy to significantly increase entropy, since $S\propto
n^{-2/3}$), and the need to raise the entropy of the more diffuse
intragroup medium.

Nevertheless, there are certainly still significant problems with our
simulations as-is, most notably the excess of late-time star formation in
the central galaxy that may be responsible for poor fits to observed X-ray
profiles.  There is also the issue of the observed scatter in the X-ray
scaling relations, which points to some more stochastic form of energy
injection.  Finally, it appears that large galaxy clusters also have
elevated entropies all the way out to their outer regions~\citep{dun08}.
Since wind energy cannot be responsible because it is less important in
higher mass systems, this suggests that AGN energy, perhaps transported
via bouyant bubbles, are needed to explain the thermodynamics of clusters.

A complete understanding of hot diffuse gas in groups and clusters will
only come when a model addresses all three ICM issues simultaneously,
namely the cooling flow, entropy, and metallicity problems.  Our current
simulations explicitly do not do so, focusing on two of the three problems
solely in the poor group regime.  Hence although our simulations suggest
that entropy injection is tied mostly to metal enrichment and not cooling
flow prevention, we forward this interpretation only a suggestion and
not a definitive statement.  This must still be tested using simulations
that self-consistently incorporate some mechanism to quench star formation
in massive galaxies in accord with cluster galaxy data.  We are actively
pursuing such investigations.

There are two immediate paths for future investigations.  The first
is to incorporate some physical or heuristic model of star formation
truncation in massive galaxies, in order to understand its impact on
energy and metal input into intragroup gas.  The second is to study
how the enrichment and entropy of intragroup gas grow with redshift
within a hierarchical framework, in order to better understand its
physical origin.  We also plan to continue developing tools to provide
closer comparisons with observations, such as the ability to produce
simulated X-ray maps that can be analyzed alongside data.  Finally,
we are planning to run significantly larger simulations with up to
$2\times 512^3$ particles in the near future, which will allow us to
have greater dynamic range, study larger clusters, and properly assess
resolution effects.  While our initial efforts towards understanding
intragroup gas chemo- and thermodynamics from outflows seem promising,
there is much work to be done to fully understand the interplay between
galaxy formation processes and the X-ray emitting plasma in the largest
bound structures in the Universe.

 \section*{Acknowledgements}
The simulations used here were run at the National Center for
Supercomputing Applications, as well as the University of Arizona's
campus supercomputers.  We thank Mark Dickinson, Mark Fardal, Neal
Katz, Dusan Keres, Trevor Ponman, David Weinberg, and the referee for
helpful suggestions.  Support for this work was provided by NASA through
grant number HST-AR-10946 from the SPACE TELESCOPE SCIENCE INSTITUTE,
which is operated by AURA, Inc. under NASA contract NAS5-26555, as
well as through grant ATP grant NNG06GH98G.  Support for this work,
part of the Spitzer Space Telescope Theoretical Research Program, was
also provided by NASA through a contract issued by the Jet Propulsion
Laboratory, California Institute of Technology under a contract with NASA.
Support was also provided by the National Science Foundation through
grant number DMS-0619881.

\end{document}